\documentclass[aps,prl,twocolumn,superscriptaddress]{revtex4-2}

\usepackage[dvipsnames]{xcolor}

\definecolor{darkred}{rgb}{0.6,0.05,0.05}
\definecolor{darkgreen}{rgb}{0.05,0.6,0.05}
\definecolor{darkblue}{rgb}{0.05,0.05,0.6}
\usepackage[colorlinks]{hyperref}
\hypersetup{colorlinks,
linkcolor=darkred,
filecolor=darkgreen,
urlcolor=darkblue,
citecolor=darkgreen}

\usepackage{tikz}

\usepackage{titlesec}

\usepackage{mathtools}
\usepackage{amsmath}
\allowdisplaybreaks
\usepackage{ bbold }
\usepackage{amssymb}
\usepackage[normalem]{ulem}
\usepackage{cancel}
\usepackage{amsthm}
\usepackage{xfrac}
\usepackage{dsfont}
\usepackage{siunitx}
\usepackage{physics}
\usepackage{graphicx}
\usepackage{hyperref}
\usepackage[capitalise]{cleveref}
\usepackage{dcolumn}
\usepackage{bm}
\usepackage{layouts}
\usepackage{comment}
\usepackage{soul}

\newcommand{\rme}{{\rm e}}
\newcommand{\rmd}{{\rm d}}
\newcommand{\rmi}{{\rm i}}

\newcommand{\LL}{{\mathcal{L}}}

\newcommand{\JJ}{{\mathcal{J}}}

\newcommand{\suppinfofootnote}{
  \footnote{For more details, see the Supplementary Information.}
}
\newcommand{\suppinfo}{\footnotemark[\value{footnote}]}

\renewcommand{\paragraph}[1]{\textit{#1}---.}

\begin{document}

\author{Filippo Ferrari}
\email{filippo.ferrari@epfl.ch}
\affiliation{Laboratory of Theoretical Physics of Nanosystems (LTPN), Institute of Physics, \'{E}cole Polytechnique F\'{e}d\'{e}rale de Lausanne (EPFL), 1015 Lausanne, Switzerland}
\affiliation{Center for Quantum Science and Engineering, \\ \'{E}cole Polytechnique F\'{e}d\'{e}rale de Lausanne (EPFL), CH-1015 Lausanne, Switzerland}
\author{Joachim Cohen}
\affiliation{Alice \& Bob, 53 boulevard du G\'en\'eral Martial Valin, 75015 Paris, France}
\author{Vincenzo Savona}
\affiliation{Laboratory of Theoretical Physics of Nanosystems (LTPN), Institute of Physics, \'{E}cole Polytechnique F\'{e}d\'{e}rale de Lausanne (EPFL), 1015 Lausanne, Switzerland}
\affiliation{Center for Quantum Science and Engineering, \\ \'{E}cole Polytechnique F\'{e}d\'{e}rale de Lausanne (EPFL), CH-1015 Lausanne, Switzerland}
\author{Fabrizio Minganti}
\email{fabrizio.minganti@gmail.com}
\affiliation{Alice \& Bob, 53 boulevard du G\'en\'eral Martial Valin, 75015 Paris, France}

\title{Bit flips are erasures in dissipative cat qubits}

\date{\today}

\begin{abstract}
Autonomous quantum error correction (QEC) stabilizes a logical manifold through dissipative events that emit into output channels, which are typically accessible to measurement. These signals are often discarded, and whether they contain useful information about logical failures remains generally unclear. Using quantum trajectories, we show that in dissipatively stabilized cat qubits bit flips are not silent logical errors: each flip is accompanied by a strong, time-localized photon burst from the dissipative buffer. Photon counting and homodyne monitoring can therefore herald the loss of logical information without interrupting the autonomous stabilization: bit flips in dissipative cat qubits are erasures. More broadly, our results show that the emitted signals of engineered reservoirs can act as built-in failure monitors for autonomous QEC, turning rare logical faults into erasures available to a decoder and reducing fault-tolerance overhead. To this end, we develop a general framework, based on past quantum states and number-resolved master equations, to quantify the detectability of such logical failures in autonomous QEC from the emitted signal.
\end{abstract}

\maketitle

\paragraph{Introduction}
Dissipation induces errors that hinder the performance of quantum computers~\cite{nielsen_quantum_2012, gottesman2009introductionquantumerrorcorrection}.
To prevent the degradation of quantum information, conventional quantum error correction (QEC) strategies repeatedly measure error syndromes and apply classical feedback~\cite{nielsen_quantum_2012,gottesman2009introductionquantumerrorcorrection, knill_theory_1997, knill_theory_2000} as done in, e.g., repetition~\cite{putterman_hardware-efficient_2025} and surface codes~\cite{dennis_topological_2002, fowler_surface_2012, krinner_realizing_2022, google_quantum_ai_suppressing_2023, google_quantum_ai_and_collaborators_quantum_2025}. 
QEC can also be performed by engineered dissipation which autonomously and continuously steers the system towards the logical code space~\cite{poyatos_quantum_1996, verstraete_quantum_2009, pastawski_quantum_2011,cohenjPRA2014, kapit_hardware-efficient_2016, reiter_dissipative_2017, lihm_implementation-independent_2018}.
Reservoir engineering protocols for QEC and state preparation are realized by nonlinearly coupling the logical system to a dissipative auxiliary mode~\cite{harrington_engineered_2022}, and have been experimentally demonstrated in, e.g., trapped ions~\cite{lin_dissipative_2013, cole_resource-efficient_2022} and circuit quantum electrodynamics (QED)~\cite{shankar_autonomously_2013, gertler_protecting_2021, li_autonomous_2024}. 

\begin{figure}
    \centering
    \includegraphics[width=0.975\linewidth]{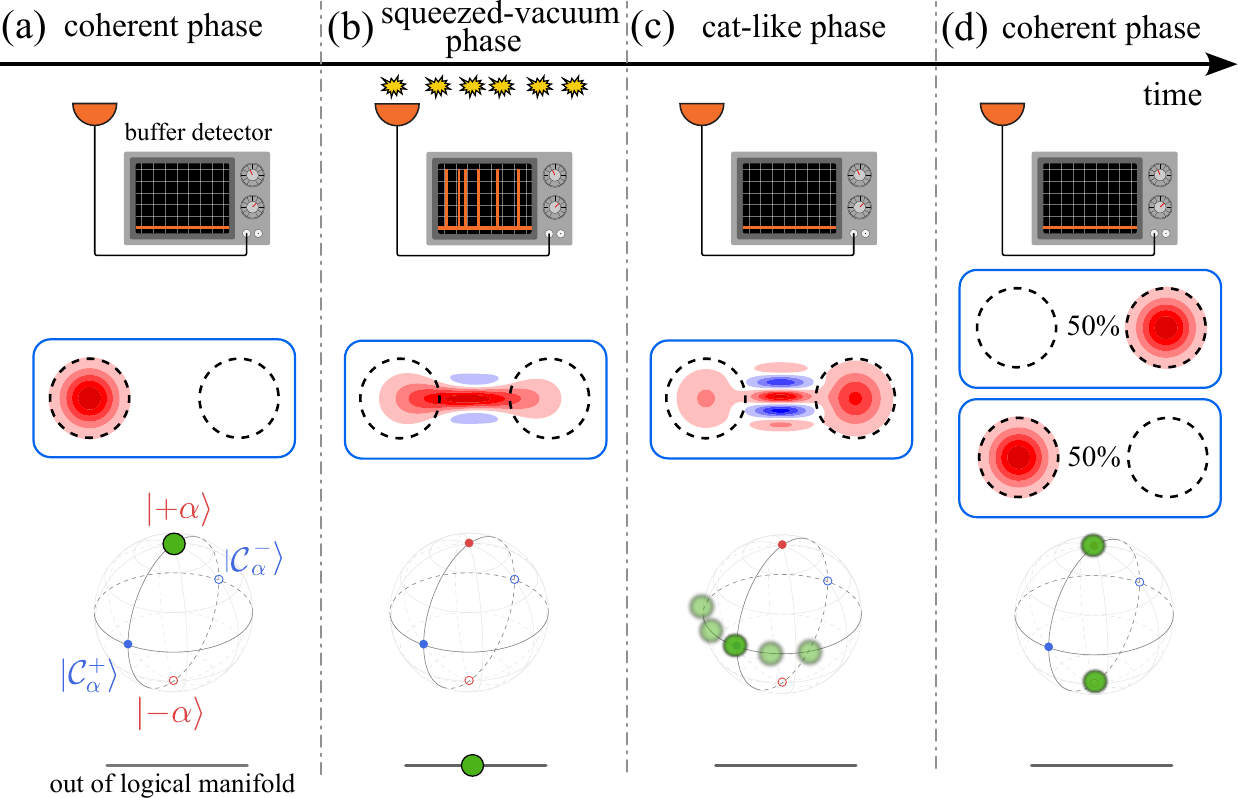}
    \caption{As a function of time and in a quantum trajectory picture, properties of bit flips making them erasures.
    A bit flip undergoes four distinct phases: 
    (a) Coherent; The buffer does not emit (lack of signal), the memory is in $\ket{+\alpha}$ (Wigner function in the middle), and the system occupies $\ket{0_L}$ in the Bloch sphere.
    (b) Squeezed-vacuum; Several buffer jumps (yellow clicks and spikes in the oscilloscope) signal the flow of the memory state towards a squeezed-like vacuum. This state is outside the Bloch sphere.
    (c) Cat-like; Once the system is re-stabilized on the logical manifold the buffer stops emitting; the state, initially an even cat, drifts on the equatorial plane of the Bloch sphere as shown by Wigner negativities of the memory.
    (d) Coherent; The buffer still shows no emission, as the system progressively loses its fringes and ends up in $\ket{\pm\alpha}$ with equal probability. 
    This description remains valid upon the more formal analysis below.
    }
    \label{fig:scheme}
\end{figure}

A paradigmatic example of autonomous QEC is the cat qubit, which encodes a qubit in a single bosonic mode rather than across many physical degrees of freedom.  
Dissipative cat qubits are generated by a tailored two-photon exchange between a high-Q mode (the memory) and a dissipative one (the buffer), stabilizing coherent states of opposite phase---the logical states $\ket{0_L}$ and $\ket{1_L}$---and their
quantum superpositions~\cite{mirrahimi_dynamically_2014, leghtas_confining_2015, lescanne_exponential_2020, grimm_stabilization_2020, reglade_quantum_2024}. 
As these coherent states are resilient to noise and unwanted dissipation, cat qubits are \textit{biased-noise qubits}: the bit-flip rate is suppressed exponentially with the cat size (the number of photons in the memory), at the price of a linear increase of the phase-flip rate.

Whether bit flips are completely suppressed--making a phase-flip repetition code sufficient to achieve fault tolerance--or residual bit flips are corrected by tailored shallow surface codes~\cite{bonilla_ataides_xzzx_2021}, cat qubits reduce the hardware footprints for fault tolerant quantum computing~\cite{guillaud_repetition_2019, putterman_hardware-efficient_2025}. 
In either case, residual or undetected bit flips can profoundly affect the performance of QEC protocols. 
Knowing the time and position of bit flips in these codes would further enhance their performance.
A heralded error that carries a classical \textit{flag} that identifies its location in space and time is known as an \textit{erasure}~\cite{grassl_codes_1997, wu_erasure_2022, teoh_dual-rail_2023, kubica_erasure_2023}.
More generally, erasure errors allow more efficient decoding and higher thresholds than generic errors: for a distance-$d$ repetition code, the number of correctable errors rises from $\lfloor(d-1)/2\rfloor$ to $d-1$. 

In this article, we discover that \textit{bit flips are erasures in dissipative cat qubits}, flagged by a strong buffer emission.
Using quantum trajectories, we show that bit flips in dissipative cat qubits are non-instantaneous events signaled by an otherwise empty buffer that starts emitting (see Fig.~\ref{fig:scheme} for a sketch),
heralding when and where a logical error occurred at no cost to the autonomous correction itself. 
Pivotal for this property is the fact that the dissipation performing autonomous QEC flows through a quasi-classical channel~\cite{mirrahimi_dynamically_2014, li_autonomous_2024}---the buffer loss, accessible for instance through its coupling to a transmission line. This property should thus be applicable to other autonoumous QEC schemes where the output channel can be monitored. 

Whether generic autonomously stabilized quantum codes yield a clean erasure flag
requires a dedicated framework which we also develop in this work.
In particular, we resort to the \textit{past quantum state} formalism~\cite{gammelmark_past_2013} and a \textit{number-resolved master equation}~\cite{levitov_electron_1996, bagrets_full_2003, flindt_full_2005, landi_current_2024, menczel_full_2026}.
We show that this answer 
depends on the encoding and the jump operator structure that performs the correction. 
For cat qubits, we demonstrate that the classical flag associated with bit flips can be detected for realistic parameters, opening an avenue for enhancing cat-qubit performance in shallow surface and repetition codes.

\paragraph{Dissipative cat qubits}
We consider two coupled driven-dissipative oscillators with bosonic annihilation operators $\hat{a}$ for the memory and $\hat{b}$ for the buffer.
The density matrix $\hat{\rho}$ evolves under the action of the Lindblad master equation~\cite{guillaud_quantum_2023}
\begin{equation}\label{eqs:two_cat_qubit}
\begin{split}
    \hat{H} = \,& g_2\left(\hat{a}^{\dagger 2}\hat{b} + \hat{a}^2\hat{b}^\dagger\right)  + \left( \varepsilon_b \, \hat{b}^\dagger + \varepsilon_b^* \,\hat{b}\right)\\
    \frac{\partial\hat{\rho}}{\partial t} = & -\rmi[\hat{H}, \hat{\rho}] + \kappa_b\mathcal{D}[\hat{b}]\hat{\rho}  + \kappa_a\mathcal{D}[\hat{a}]\hat{\rho} + \kappa_\phi\mathcal{D}[\hat{a}^\dagger\hat{a}]\hat{\rho},
\end{split}
\end{equation}
where $\hat{H}$ describes the parametric down-conversion exchange at a rate $g_2$ between memory and buffer and a coherent single-photon drive with amplitude $|\varepsilon_b|$ on the buffer.
We include in the model the following incoherent processes: buffer (memory) single-photon losses at a rate $\kappa_a$ ($\kappa_b$) and dephasing on the memory at a rate $\kappa_\phi$, acting through the Lindblad dissipators $\mathcal{D}[\hat{L}]\hat{\rho} = \hat{L} \hat{\rho} \hat{L}^\dagger -  \{\hat{L}^\dagger \hat{L}, \hat{\rho}\} /2$.

\begin{figure}
    \centering
    \includegraphics[width=1\linewidth]{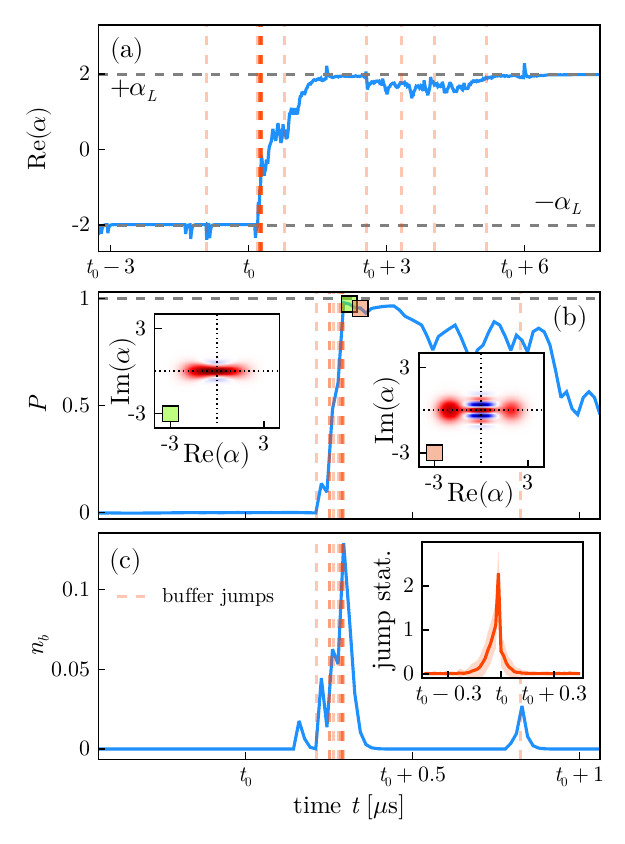}
    \caption{Quantum trajectory analysis of a bit flip event as a function of time. 
    (a) Evolution of $\alpha = \expval{\hat{a}}{\Psi(t)}$. Vertical dashed lines indicate emission of photons from the buffer.
    (b) Evolution of $P = \expval{\hat{P}}{\Psi(t)}$ at the initial bit-flip times. 
    Insets: memory Wigner functions at times indicated by green and sienna markers in the main panel.
    (c) The photon number in the buffer $n_b = \expval{\hat{b}^\dagger \hat{b}}{\Psi(t)}$ around the initial time of the bit flip. 
    Inset: statistics of the buffer emission over $10^3$ trajectories (solid line: average; dashed line: variance) showing an enhanced emission at the beginning of a bit flip event.
    The parameters are $\kappa_a/2\pi = \kappa_\phi/2\pi=10$ kHz, $g_2/2\pi = 5$ MHz, $\kappa_b/2\pi= 100$ MHz. 
    We fix $\varepsilon_b$ so that the memory hosts $4$ photons in the steady state.
    }
    \label{fig:quantum_trajectories}
\end{figure}

\paragraph{Quantum trajectories}
In the absence of photon loss and dephasing, the system hosts stable coherent states $\ket{\pm\alpha}$ of opposite phase defining the north and south pole of the logical Bloch sphere, as well as their quantum superpositions~\cite{mirrahimi_dynamically_2014}.
Bit-flips are due to the presence of these dissipative terms and can be visualized using quantum trajectories, i.e., wave functions evolving smoothly under a non-Hermitian effective Hamiltonian that is stochastically interrupted by quantum jumps, representing photon emission and dephasing events~\cite{dalibard_wave-function_1992, molmer_monte_1993, daley_quantum_2014}.
Averaging many trajectories reproduces the density matrix evolution.
Details on quantum trajectories to describe bit- and phase-flips in cat states can be found in, e.g., Refs.~\cite{bartolo_homodyne_2017, Minganti_2018, ferrari2026bitflipssaturationquantum}.

We show the bit flip of a representative quantum trajectory in Fig.~\ref{fig:quantum_trajectories}.
First, we focus on the memory state.
We plot $\textrm{Re}(\alpha)$ with $\alpha=\bra{\Psi(t)}\hat{a}\ket{\Psi(t)}$ in Fig.~\ref{fig:quantum_trajectories} (a).
The bit flip occurs in a typical time of the order of $1/[\kappa_a\langle\hat{a}^\dagger\hat{a}\rangle]$. 
There is a long transient during which $|\alpha| < |\alpha_L|$ (dashed horizontal lines), with $|\alpha_L|^2 \simeq n_a$ the photon number in the steady state.
We conclude that the state is not coherent during the bit flip.
We plot the parity of the memory $\hat{P} = \exp(\rmi \pi \hat{a}^\dagger \hat{a})$ at the initial time of the bit flip in Fig.~\ref{fig:quantum_trajectories} (b). 
The initial coherent state $\ket{\Psi(t)} \simeq \ket{\alpha}$ shrinks to a squeezed vacuum with $P = \bra{\Psi(t)}\hat{P}\ket{\Psi(t)} \simeq +1$, as also shown by the Wigner function in the inset.
Then, the even cat is stabilized, as marked by both $\hat{P}$ and the Wigner function in the second inset, and the parity is progressively degraded by the action of single-photon losses.
As sketched in Fig.~\ref{fig:scheme}, we can thus distinguish four phases: coherent; squeezed vacuum; cat-like; coherent again~\suppinfofootnote\,(see also Ref.~\cite{ferrari2026bitflipssaturationquantum}).

We now investigate the buffer dynamics, whose emission concentrates in the initial stage of the bit flip (red-dashed lines in all the panels of Fig.~\ref{fig:quantum_trajectories}).
This emission is centered around a time $t_0$, where the memory is in the squeezed-vacuum-like phase and its population is minimal.
At this time, the buffer population significantly deviates from zero [cf. Fig.~\ref{fig:quantum_trajectories} (c)].
The analysis over $10^3$ trajectories exhibiting a bit flip, reported in the inset of Fig.~\ref{fig:quantum_trajectories} (c), confirms that a strong buffer emission signals the beginning of a bit-flip event. 

Overall, the quantum trajectory analysis paints a clear picture: the bit flip occurs when the memory state flows to a squeezed vacuum (a state with even parity) and then gets quickly restabilized into the even cat state.
This is accompanied by a strong and time-localized emission of the buffer, a classical flag making it possible to know if such an event has occurred and suggesting that bit-flip errors are erasures.
The remainder of the paper is dedicated to formalizing this intuition.

\paragraph{Past quantum states}
Quantum trajectories require an explicit modeling of how the signal escaping from buffer and memory is collected, i.e., the so-called unraveling~\cite{breuer_theory_2007}.
Furthermore, since bit flips are rare events, it is computationally expensive to accumulate sufficient statistics for the calculation of average quantities during a bit flip.
Thus, we employ a master equation framework directly yielding such values.

We consider a system initialized at time $t=0$ in one logical state, $\hat{\rho}^i=\hat{\rho}^{+z}$, and postselected at a later time $T$ in the opposite logical sector, described by the projector $\hat{\Pi}^{-z}$.
Notice that $\hat{\Pi}^{\pm z}$ is the dual of $\hat{\rho}^{\pm z}$, acting as a syndrome-plus-decoder that assigns each state to a pole's basin of attraction~\suppinfo.
This pre- and postselected evolution is described by the past quantum state $\Xi(t)=\{\hat{\rho}(t),\hat{E}(t)\}$~\cite{gammelmark_past_2013}. 
The state $\hat{\rho}(t)$ propagates the initial state $\hat{\rho}^i$ forward in time up to $t$. 
The effect matrix $\hat{E}(t)$ conditions observables on finding the system in the final logical sector $\hat{\Pi}^{f}=\hat{\Pi}^{-z}$, and is obtained by propagating $\hat{\Pi}^{f}$ backward from the final time $T$ to the earlier time $t$. They obey
\begin{equation}\label{eqs:past_quantum_state_definition}
\hat{\rho}(t)=\rme^{\mathcal{L}t}\hat{\rho}^i, \qquad \hat{E}(t)=\rme^{\mathcal{L}^{\dagger}(T-t)}\hat{\Pi}^{f}.
\end{equation}
Here $\mathcal{L}$ is the
Liouvillian generating the Lindblad master equation, and
$\mathcal{L}^{\dagger}$ is its adjoint.  The logical projectors satisfy
$\operatorname{Tr}[\hat{\rho}^{\pm z}\hat{\Pi}^{\pm z}]=1$ and
$\operatorname{Tr}[\hat{\rho}^{\pm z}\hat{\Pi}^{\mp z}]=0$, with
$\hat{\sigma}^{z}_{L}=\hat{\Pi}^{+z}-\hat{\Pi}^{-z}$.
The buffer population and the total number of photons emitted, conditioned on the occurrence of a bit flip, are  
\begin{equation}\label{eqs:past_quantum_state_emission}
    n_{\rm out}(t)
    =\kappa_b
    \frac{
        \operatorname{Tr}\!\left[
            \hat{E}(t)\,\hat{b}^{\dagger}\hat{b}\,\hat{\rho}(t)
        \right]
    }{
        \operatorname{Tr}\!\left[
            \hat{E}(t)\hat{\rho}(t)
        \right]
    } , \quad \mathcal{E}
    = \int_0^T \rmd t\, n_{\rm out}(t).
\end{equation}
The same construction can be used for the reverse flip by exchanging
$\hat{\rho}^{+z}\leftrightarrow\hat{\rho}^{-z}$ and
$\hat{\Pi}^{-z}\leftrightarrow\hat{\Pi}^{+z}$.

\begin{figure}
    \centering
    \includegraphics[width=\linewidth]{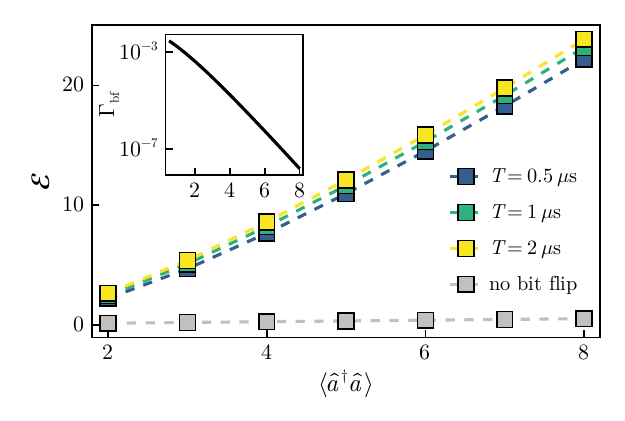}
    \caption{Conditioned buffer emission $\mathcal{E}$ given by Eq.~\eqref{eqs:past_quantum_state_emission} as a function of $\langle\hat{a}^\dagger\hat{a}\rangle$. 
    The gray markers correspond to $i=f=+z$ in Eq.~\eqref{eqs:past_quantum_state_definition} (no bit flip) for $T=1\,\mu\textrm{s}$. 
    The colored markers correspond to $i=+z$, $f=-z$ (a bit flip has occurred) for $T=0.5\,\mu\textrm{s}$ (blue), $T=1\,\mu\textrm{s}$ (green), and $T=2\,\mu\textrm{s}$ (yellow).
    Here, $\kappa_a/2\pi=\kappa_\phi/2\pi=5\,\textrm{kHz}$.
    The inset shows the bit-flip error rate $\Gamma_{\rm bf}$ of the system. 
    Other parameters as in Fig.~\ref{fig:quantum_trajectories}.
    }
    \label{fig:past_quantum_state}
\end{figure}

We plot $\mathcal{E}$ as a function of the cat size,
$\langle \hat{a}^{\dagger}\hat{a}\rangle
=
\operatorname{Tr}[
\hat{\rho}_{\rm ss}\hat{a}^{\dagger}\hat{a}]$,
with $\hat{\rho}_{\rm ss}$ the system steady state, and for different
postselection times $T$ in Fig.~\ref{fig:past_quantum_state}.  Without a bit
flip, the emitted photon number remains close to zero and marginally increases with cat size.  
Conditioned on a bit flip, instead, the buffer emits a much larger number of photons, which grows with cat size and is only weakly affected by $T$.  
This confirms the trajectory picture: bit flips are heralded by a strong, rapid, and time-localized buffer emission.
The large contrast between the flip and no-flip cases indicates that this
emission provides an experimentally detectable erasure flag, independently of
the detection protocol.
In the inset we plot the bit-flip error rate $\Gamma_{\rm bf}$ which decreases exponentially with the cat size $\langle\hat{a}^\dagger\hat{a}\rangle$, indicating autonomous QEC.

\begin{figure}
    \centering
    \includegraphics[width=\linewidth]{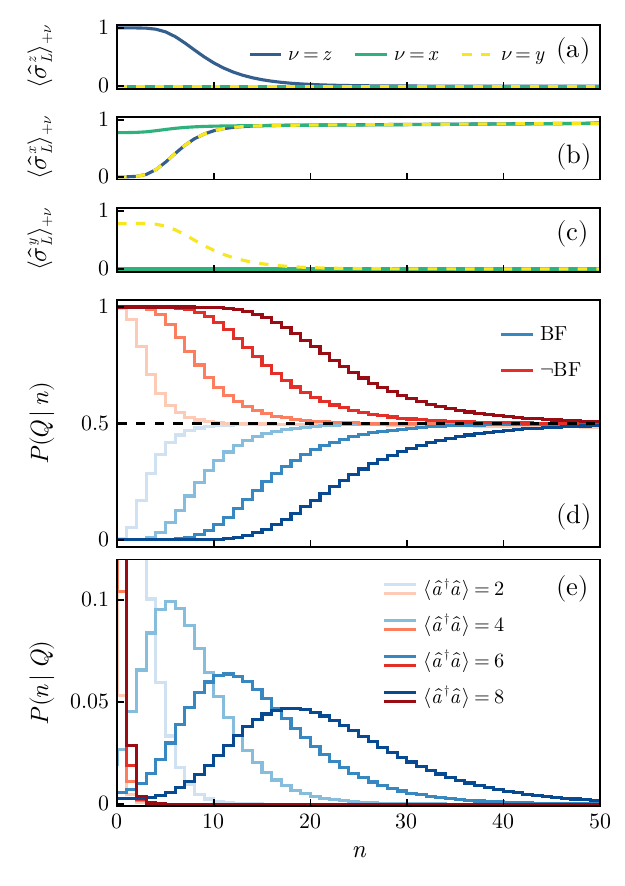}
    \caption{Number-resolved master equation results. 
    (a-c) Expectation value of the logical Pauli operator, $\langle\hat{\sigma}_L^\mu(T, n)\rangle_{+\nu}$ corresponding to $\hat{\rho}(t=0)=\hat{\rho}^{+\nu}$ with $\nu=\{x, y, z\}$ (blue, green and yellow lines respectively), for (a) $\mu=z$, (b) $\mu=x$, (c) $\mu=y$, as a function of $n$.
    The memory photon number is $\langle\hat{a}^\dagger\hat{a}\rangle=4$.
    (d) Probability $P(Q|n)$ of event $Q\in\{\textrm{BF (blue)},\neg\textrm{BF  (red)} \}$, conditioned on the occurrence of $n$ jumps as a function of $n$ for $\langle\hat{a}^\dagger\hat{a}\rangle=2, 4, 6, 8$ (from light to dark color).
    (e) Probability $P(n|Q)$ of having $n$ jumps given the event $Q$ as a function of $n$.
    Color scheme as in panel (d).
    $T$ is fixed to $T=1\,\mu\textrm{s}$.
    Parameters as in Fig.~\ref{fig:past_quantum_state}.
    }
    \label{fig:full_counting_statistics}
\end{figure}

\paragraph{Number-resolved master equation} 
We want to use the buffer emission to flag the loss of quantum information.
For this task, we must model how the signal of the buffer is collected. We assume an ideal photodetector clicking every time the buffer emits a photon and write $\hat{\rho}(t)=\sum_n\hat{\rho}_n(t)$, where $\hat{\rho}_n(t)$ is the unnormalized density matrix conditioned on the detection of $n$ buffer photons~\cite{menczel_full_2026}.
The probability of observing $n$ buffer jumps is $P_t(n) = \operatorname{Tr}[\hat{\rho}_n(t)]$.
The $\hat{\rho}_n(t)$ evolve with a number-resolved master equation~\suppinfo.

First, we investigate what happens to the expectation value of the logical Pauli operators conditioned on the occurrence of $n$ buffer jumps.
This is given by $\langle \hat{\sigma}_L^{\mu}(T, n) \rangle_{\pm\nu} = \operatorname{Tr}\left[\hat{\rho}^{\pm\nu}_{n}(T)\,\hat{\sigma}_L^{\mu}\right]/\operatorname{Tr}[\hat{\rho}^{\pm\nu}_{n}(T)]$ with $\mu,\nu=\{x, y, z\}$, where $\hat{\sigma}_L^{\mu} = \hat{\Pi}^{+\mu}-\hat{\Pi}^{-\mu}$ and $\hat{\rho}^{\pm\nu}_{n}(T) \equiv \hat{\rho}_{n}(T)$ for $\hat{\rho}(t=0) = \hat{\rho}^{\pm\nu}$~\suppinfo. 
In Figs.~\ref{fig:full_counting_statistics} (a-c), we plot $\langle \hat{\sigma}_L^{\mu}(T, n) \rangle_{+\nu}$ for the logical states $\hat{\rho}(t=0) = \hat{\rho}^{+\nu}$ [results for $\hat{\rho}(t=0) = \hat{\rho}^{-\nu}$ are reported in the Supplementary Information].
Upon the occurrence of a sufficient number of jumps, all initial states collapse approximately to the $\hat{\rho}^{+x}$ state, confirming the prediction of quantum trajectory analysis in Fig.~\ref{fig:quantum_trajectories}.

Fig.~\ref{fig:full_counting_statistics} (d) shows $P(Q|n) = \operatorname{Tr}[\hat{\Pi}^{f} \hat{\rho}^{+ z}_{n}(T)]/\operatorname{Tr}[\hat{\rho}^{+ z}_{n}(T)]$, where $Q\in\{{\rm BF}, \neg {\rm BF}\}$. 
For our choice of initial state, $P({\rm BF}|n)$ is the bit-flip probability upon the occurrence of $n$ jumps, i.e.,  measuring $1_L$ ($\hat{\Pi}^{f} =\hat{\Pi}^{-z}$), having initialized the system in $0_L$.
Similarly, $P(\neg {\rm BF}|n)$ is the probability of no-bit-flips.
As expected, the probability converges to $P(Q|n) = 1/2$, indicating that $\hat{\rho}^{+ z}_{n}(T) \propto \hat{\rho}^{+x}$.
Using Bayes' theorem, we obtain the full counting probability distribution conditioned on $Q$,
$P(n|Q) = {P(Q|n) \,  P(n)}/{P(Q)}$, with $P(n) = \operatorname{Tr}[\hat{\rho}_n(T)]$, and $P(Q) =  \operatorname{Tr}[\hat{\Pi}^{f} \hat{\rho}^{+ z}(T)]$ representing the probability of a jump for $\hat{\rho}^{+ z}(T)=\rme^{\LL T}\hat{\rho}^{+ z}$, unconditioned to the number of jumps.
Results are reported in Fig.~\ref{fig:full_counting_statistics} (e).
The distribution $P(n|1_L)$ has little overlap with the distribution $P(n|0_L)$ with this overlap decreasing with larger memory photon number $\langle\hat{a}^\dagger\hat{a}\rangle$.

In summary, we have demonstrated that, upon the occurrence of a sufficient number of jumps, any state on the Bloch sphere converges toward $\hat{\rho}^{+x}$, resulting in a complete loss of quantum information.
We conclude that the initial dynamics of what, in a coarse-grained picture, appears as bit-flip events can be modeled not as the action of Pauli $x$ operators, but rather as $\hat{J}_{\rm BF} \simeq \ketbra{\mathcal{C}_\alpha^+}{\mathcal{C}_\alpha^-}$, with $\ket{\mathcal{C}_\alpha^\pm} \propto \ket{\alpha} \pm \ket{-\alpha}$ representing the cat states, followed by the loss of parity due to memory single-photon jumps.
Despite the slow convergence of the state with $n$, the bimodal nature of the emission makes it possible to apply a threshold to the signal thereby minimizing errors in detecting bit flips in a monitored system.

\paragraph{Threshold for the buffer emission}
We define the event $E$ as any process associated with the loss of quantum information.
To quantify the loss of information for states along the logical $z$ basis, we
compare the output states obtained from $\hat{\rho}^{+z}_{n}(T)$ and $\hat{\rho}_{n}^{-z}(T)$. We define the
remaining $z$-basis distinguishability and the minimum probability of incorrectly inferring the initial $z$ state given $n$ jumps as
\begin{equation}\label{eqs:probability_quantum_info}
\begin{split}
&D_n^z = \left\|\frac{\hat{\rho}^{+z}_{n}(T)}{\operatorname{Tr}[\hat{\rho}^{+z}_{n}(T)]}-\frac{\hat{\rho}^{-z}_{n}(T)}{\operatorname{Tr}[\hat{\rho}^{-z}_{n}(T)]}\right\|_1,\\& 
P(E|n) = 1-\frac{D_n^z}{2},
\end{split}
\end{equation}
being $||\cdot||_1$ the trace distance.
We seek a simple threshold criterion based on the number of emitted photons $\bar{n}$. 
We thus assign a cost $C_{\mathrm{FP}}$ to a false positive detection, corresponding to declaring $E$ when the event did not occur.
A false negative event, corresponding to declaring $\neg E$ when the event did occur, has cost
$C_{\mathrm{FN}} = \zeta \, C_{\mathrm{FP}}$, where $\zeta$ is a proportionality factor to be determined by the specifics of the error correction code under consideration.
The conditional expected cost of declaring $E$, given the observed value $n$, is
then 
$\mathcal{C}_{\mathrm{FP}}(n)=C_{\mathrm{FP}} P(\neg E \mid n)=
C_{\mathrm{FP}}\left[1-P(E\mid n)\right].
$
Similarly, the conditional expected cost of declaring no event is
$
\mathcal{C}_{\mathrm{FN}}(n)
=
C_{\mathrm{FN}} P(E\mid n) = \zeta \, C_{\mathrm{FP}} \, P(E\mid n)
$.

\begin{figure}
    \centering
    \includegraphics[width=\linewidth]{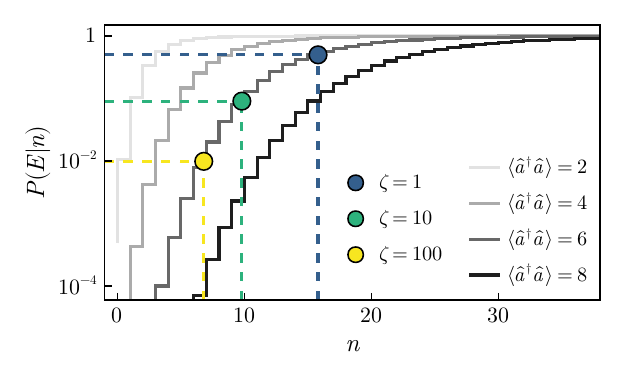}
    \caption{As a function of $n$, probability $P(E|n)$ that quantum information has been erased given $n$ buffer jumps, as in Eqs.~\eqref{eqs:probability_quantum_info}.
    Solid lines refer to different mean memory photon numbers. 
    The markers point to the value of $\overline{n}$ obtained according to Eq.~\eqref{eq:n_bar} for different values of $\zeta$ and for $\langle\hat{a}^\dagger\hat{a}\rangle=6$.
    Parameters as in Fig.~\ref{fig:past_quantum_state}.
    }
    \label{fig:threshold_probabilities}
\end{figure}

We then declare $E$ whenever
$\mathcal{C}_{\mathrm{FP}}(n) < \mathcal{C}_{\mathrm{FN}}(n)$
and therefore we introduce the threshold
\begin{equation}\label{eq:n_bar}
\bar{n}= \min \left\{ n \text{ such that } P(E\mid n)
>
\frac{1}
{1 + \zeta} \right\}.
\end{equation}
In Fig.~\ref{fig:threshold_probabilities} we plot $P(E|n)$ for various $\langle\hat{a}^\dagger\hat{a}\rangle$ and as a function of $n$.
The threshold $\overline{n}$ is indicated by the vertical lines intersecting $P(E|n)$ for $\langle\hat{a}^\dagger\hat{a}\rangle=6$.
Typically, the cost of a missed bit flip is larger than the cost of a false alarm. We thus consider $\zeta\in [1, 10, 100]$.
The line where $P(E\mid n)$ intersects that of $\zeta$ indicates the ratio of false-positive rejections. 
Note that curves with increasing $\zeta$ are fairly close to each others, showing that even for moderate values of $\zeta$ the protocol is advantageous. 
Systems with larger cat sizes, resulting in lower bit-flip rates, will favor larger values of $\bar{n}$ to get the same false-positive rejection rate, in line also with the decreasing probability of observing a bit flip (as captured by $\Gamma_{\rm bf}$ obtained by studying the Liouvillian gap in the inset of Fig.~\ref{fig:past_quantum_state}).

\paragraph{Imperfect photodetectors and homodyne detection}
As we show in the End Matter, including realistic values for detector efficiencies and dark counts does not qualitatively change the previous results.
Although remarkable progress has been made in building efficient detectors for microwave photons in circuit QED~\cite{chen_microwave_2011, besse_single-shot_2018, kono_quantum_2018, dassonneville_number-resolved_2020, lescanne_irreversible_2020}, homodyne detection of the output signal remains a more common and accessible form of measurement.
We briefly comment on the possibility of detecting bit flips using homodyne detection in the End Matter, thereby exploiting the fact that the output signal emitted is mostly coherent.

\paragraph{Erasures in autonomous QEC}
Having shown the erasure-like nature of bit flips in the cat code, a natural question is how general the properties found here are.
As we have shown, in the autonomous stabilization of cat qubits multiple buffer jumps (detectable clicks) are required to steer the system back to the code manifold. 
Conversely, single click does not bring the state back deterministically to the code space because the engineered dissipation is ``local'' in phase space.
To test whether these features are necessary for other autonomous codes, we consider examples of autonomously stabilized quantum codes in the End Matter and Supplementary Information. We confirm that these three properties are necessary, while also demonstrating the generality of the methods introduced here for characterizing the features of bit flips.

\paragraph{Conclusion}
We have demonstrated that bit-flip events are associated with large buffer emission.
Conversely, we have shown that the emission of photons from the buffer can be used as a classical flag to signal the loss of quantum information.
We proved our statement using multiple open-system approaches, specifically i) quantum trajectories, ii) past quantum state formalism, and iii) number-resolved master equations.
We showed the viability of the protocol considering a simple criterion to choose the threshold based on the total emission.
What makes this mechanism particularly exploitable is that it comes essentially ``for free": while erasure detection in other architectures requires monitoring the system itself~\cite{wu_erasure_2022, teoh_dual-rail_2023, ma_high-fidelity_2023, koottandavida_erasure_2024, levine_demonstrating_2024, shaw_erasure_2025, de_graaf_mid-circuit_2025, zhang_logical_2026}, here the stabilization mechanism heralds its own failure.
Detecting an erasure thus amounts to collecting a signal already emitted rather than performing a dedicated syndrome measurement.

Our finding opens doors for concrete improvements to error correction protocols in bosonic codes, as well as perspective for real-time control of open quantum systems \cite{Minev2019}.
The systematic detection of bit flips should improve the capability of correcting errors in
cat-qubit devices and enhance their hardware-efficient nature.
As such, our discovery further boosts the viability of cat qubits toward fault-tolerant quantum computing.

\begin{acknowledgments}
\paragraph{Acknowledgements}
We acknowledge useful discussions with G. Campanaro, A. Essig, A. Melo, and P. Winkel.
This work was supported by the Swiss National Science Foundation through Project No. 200020\_215172.
All numerical simulations in this work were performed with the packages \texttt{QuantumToolbox.jl}~\cite{mercurio_quantum_2025}, \texttt{Qutip}~\cite{johansson_qutip_2012}, and \texttt{Dynamiqs}~\cite{guilmin2025dynamiqs}.
\end{acknowledgments}

\onecolumngrid
\begin{center}
  \textbf{\large End Matter}\\[1em]
\end{center}
\vspace{1em}
\twocolumngrid

\paragraph{Imperfect photodetector}
Contrary to the ideal photodetector studied in the main text, realistic apparatuses have: 
(i)
A finite efficiency $\eta$ that quantifies the fraction of photons that are not detected by the device;
(ii) A dark-photon count rate $\gamma_\rmd$ that accounts for detected events in the absence of photons emitted by the cavity.
Values of $\eta$ and $\gamma_\rmd$ characterizing good but not state-of-the-art devices are $\eta=70\%$ and $\gamma_\rmd/2\pi=1\,\textrm{kHz}$~\suppinfo.

\begin{figure}
    \centering
    \includegraphics[width=\linewidth]{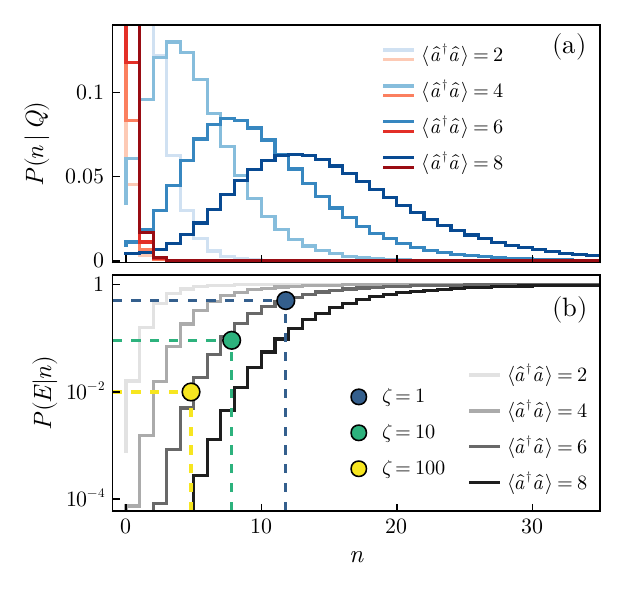}
    \caption{For an imperfect photodetector and as a function of the number $n$ of buffer jumps:
    (a) $P(n|Q)$, the probability of having $n$ buffer jumps given the event $Q \in \{{\rm BF}, \, \neg {\rm BF} \}$ [c.f. Fig.~\ref{fig:full_counting_statistics}(c) for the ideal case].
    (b) $P(E|n)$, the probability that quantum information is lost after $n$ buffer jumps; the vertical lines indicate the threshold $\bar{n}$ for various values of $\zeta$ (see Fig.~\ref{fig:threshold_probabilities} for an ideal photodetector).
    Parameters as in Fig.~\ref{fig:past_quantum_state}; efficiency $\eta=70\%$ and dark count rate $\gamma_\rmd/2\pi=1\,\textrm{kHz}$.
    }
    \label{fig:imperfect_photodetector}
\end{figure}

In Fig.~\ref{fig:imperfect_photodetector} (a) we show the full counting probability distribution conditioned on the presence/absence of a bit-flip event.
Compared to the ideal case in Fig.~\ref{fig:full_counting_statistics} (c), imperfections slightly shifts the peak of $P(n|1_L)$ towards lower $n$ and closer to $P(n|0_L)$.
The two distributions remain nevertheless well separated, and their overlap decreases with $\langle\hat{a}^\dagger\hat{a}\rangle$.
In Fig.~\ref{fig:imperfect_photodetector} (b) we plot the probability that quantum information has been lost given $n$ jumps using Eq.~\eqref{eqs:probability_quantum_info}.
Compared with Fig.~\ref{fig:threshold_probabilities}, $P(E|n)$ and $\overline{n}$ are shifted towards the left.
Overall, despite the presence of multiple non-idealities, the optimal $\overline{n}$ corresponds to a minimal overlap between $P(n|0_L)$ and $P(n|1_L)$, confirming the viability of realistic bit-flip detection with a photon counting measurement scheme.

\begin{figure}
    \centering
    \includegraphics[width=\linewidth]{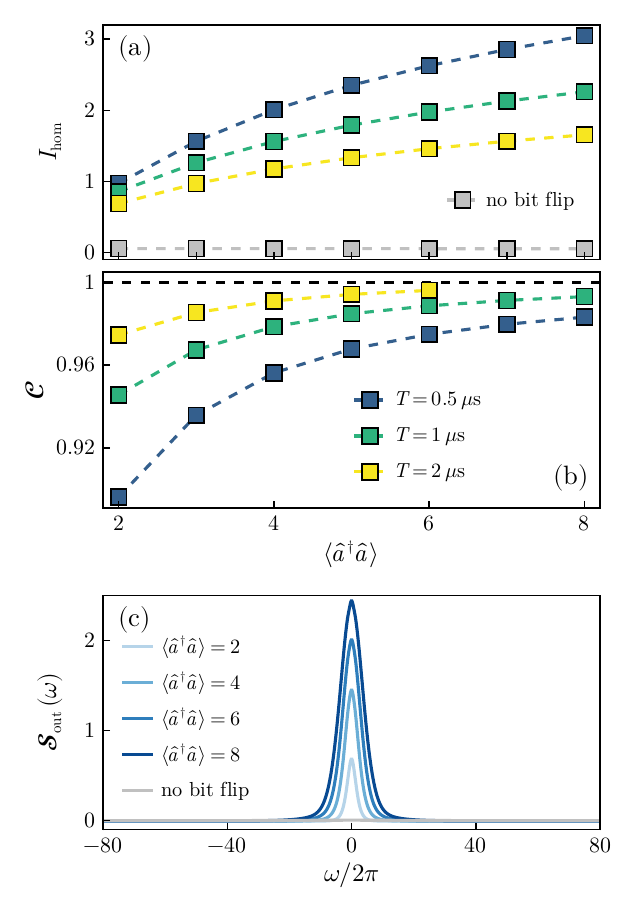}
    \caption{Homodyne detection of bit flips.
    As a function of the memory mean photon number $\langle\hat{a}^\dagger\hat{a}\rangle$ and for different integration times $T$, (a) Integrated homodyne photocurrent $I_{\rm hom}$ in Eq.~\eqref{Eq:homodyne_signal} and (b) Cauchy-Schwarz ratio $\mathcal{C}$ conditioned on the occurrence of bit flips [see Eq.~\eqref{eq:coherence}].
    (c) Conditioned emission spectrum $S_{\rm out}(\omega)$ in Eq.~\eqref{eq:emission_spectrum} for $T=1\,\mu\textrm{s}$.
    The gray line corresponds to $i=f=+z$.
    The colored lines correspond to $i=+z$, $f=-z$ for $\langle\hat{a}^\dagger\hat{a}\rangle=2, 4, 6, 8$.
    Parameters as in Fig.~\ref{fig:past_quantum_state}.
    }
    \label{fig:homodyne}
\end{figure}

\begin{figure}
    \centering
    \includegraphics[width=\linewidth]{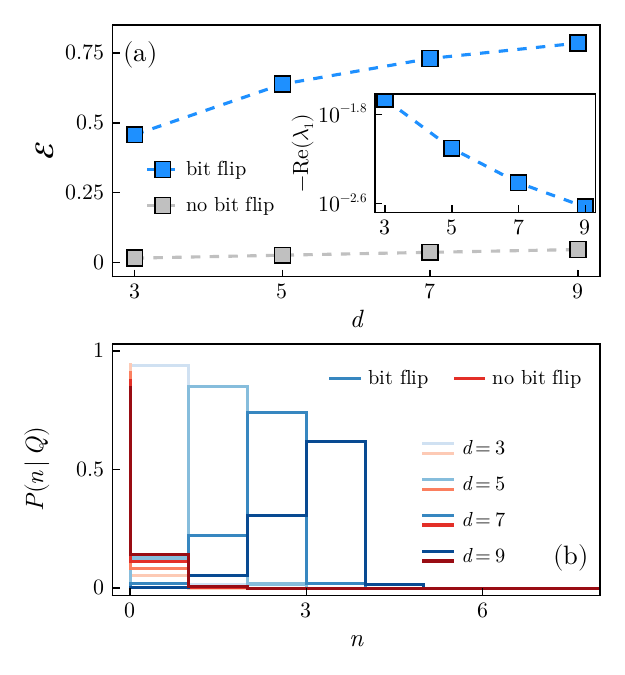}
    \caption{
    Dissipative repetition code of distance $d$. (a) Mean ``code emission'' conditioned on bit flips as a function of $d$. 
    The inset shows the bit-flip error rate $\Gamma_{\rm bf}$ of the system. 
    (b) Probability $P(n|Q)$ for $d=3, 5, 7, 9$ as a function of $n$ for $Q \in \{{\rm BF}, \, \neg {\rm BF} \}$.
    Code parameters are $\gamma=0.1$, $\kappa=1$.
    }
    \label{fig:repetition_code}
\end{figure}

\paragraph{Homodyne detection}
In circuit QED, homodyne detection is a common scheme to measure the quadratures of an oscillator. The emitted field interferes with a strong coherent reference field and the resulting integrated homodyne photocurrent (here for the buffer mode) is $J_{\rm hom}(t) = \sqrt{\kappa_b}\langle\hat{b} + \hat{b}^\dagger\rangle(t) + \xi(t)$, where $\langle\xi(t)\xi(t')\rangle=\delta(t-t')$ is a white noise term arising from quantum vacuum fluctuations and detector noise~\cite{wiseman_quantum_1993_1, wiseman_quantum_1993_2, wiseman_quantum_2009}. 
We consider the integrated homodyne signal over the time interval $T$
\begin{equation}\label{Eq:homodyne_signal}
I_{\rm hom} = \sqrt{\kappa_b/T}\int_0^{T}\rmd t\,\langle\hat{b} + \hat{b}^\dagger\rangle(t) + \mathcal{N}(0, 1),
\end{equation}
where $\mathcal{N}(0, 1)$ is a Gaussian noise with zero mean and unit variance.
To determine the homodyne photocurrent conditioned on the occurrence or absence of bit flips we employ the past quantum state formalism~\suppinfo.
We plot $I_{\rm hom}$ for various cat sizes and $T$ in Fig.~\ref{fig:homodyne} (a).
$I_{\rm hom}$ grows with $\langle\hat{a}^\dagger\hat{a}\rangle$. 
Shorter times yield more distinguishable results for the presence or absence of bit flips. Indeed, $I_{\rm hom}\simeq 0$ far from the initial part of the jump or in the absence of bit flips, and integrating over longer times simply reduces the photocurrent.

Homodyne measurements require coherent signals. 
We estimate the coherence of the buffer emission, conditioned to the presence/absence of a bit flip, using the Cauchy-Schwarz ratio
\begin{equation}\label{eq:coherence}
    \mathcal{C} = \left|\int_0^T\rmd t\,\langle\hat{b}\rangle(t)\right|^2 \Bigg/\left(T\int_0^T\rmd t\,\left|\langle\hat{b}\rangle(t)\right|^2\right).
\end{equation}
In Fig.~\ref{fig:homodyne} (b) we plot $\mathcal{C}$ as a function of cat size for various integration times $T$. 
We find $\mathcal{C}\gtrsim0.9$ if $\langle\hat{a}^\dagger\hat{a}\rangle\ge2$.
As a coherent signal yields $\mathcal{C} = 1$, we conclude that the emitted signal is highly coherent.

To determine the demodulation frequency of the output signal and the bandwidth needed to capture its components we study the buffer emission spectrum $S_{\rm out}(\omega)$.
The quantum regression theorem, conditioned on the occurrence or absence of a bit flip, gives
\begin{equation}\label{eq:emission_spectrum}
    S_{\rm out}(\omega) = \kappa_b\int_0^T\rmd t_1\int_0^T\rmd t_2\,\rme^{\rmi\omega(t_2-t_1)}G_{\rm out}^{(1)}(t_2, t_1),
\end{equation}
where the conditioned first-order coherence function is
\begin{equation}
    G_{\rm out}^{(1)} = \frac{\mathrm{Tr}\!\left[\hat{E}(t_2)\,\hat{b}^\dagger\rme^{\LL(t_2-t_1)}\,\hat{b}\,\hat{\rho}(t_1)\right]}{\mathrm{Tr}\!\left[\hat{E}(t)\,\hat{\rho}(t)\right]},
\end{equation}
and $\hat{\rho}(t_1) $ and $\hat{E}(t_2)$ are defined in Eq.~\eqref{eqs:past_quantum_state_definition}.
As shown in Fig.~\ref{fig:homodyne} (c), for no bit flips $S_{\rm out}(\omega)$ stays flat (no emitted signal).
During a bit flip, instead, $S_{\rm out}(\omega)$ displays a sharp peak whose height grows with $\langle\hat{a}^\dagger\hat{a}\rangle$, whose linewidth is much smaller than $\kappa_b$ and centered around $\omega = 0$.
That is, the signal can be collected and demodulated around the buffer drive frequency.

\paragraph{Dissipative repetition code}
We study here the bit flips of a dissipative repetition codes with features similar to the cat code.
Consider a system of $j=1,...,d$ qubits subject to the jump operators $\hat{L}_j^{\rm bf} = \sqrt{\gamma/2}\,\hat{\sigma}_j^x$.
The logical states of the code are $\ket{0_L}=\ket{0\,...\,0}$ and $\ket{1_L}=\ket{1\,...\,1}$.
The repetition code detects the syndromes $\mathbf{s}=(s_1,...,s_{d-1})$, with $s_j=\pm1$, and applies the corresponding recovery.
To create an autonomous version we translate the detection-and-recovery procedure into a set of $2^{d-1}-1$ error-correcting jump operators.
They read~\suppinfo
\begin{equation}
\begin{split}
    \hat{L}_{\bf s}^{\rm corr} &= \sqrt{\kappa}\hat{R}_{\bf s}\hat{P}_{\bf s},\\
    \hat{P}_{\bf s} = \prod_{j=1}^{d-1}\frac{\mathds{1}+s_j\hat{S}_j}{2},&\qquad\hat{R}_{\bf s} = \frac{1}{\sqrt{w_{\bf s}}}\sum_{j:\,e_j=1}\hat{\sigma}_j^x.
\end{split}
\end{equation}
$\hat{P}_{\bf s}$ project the state onto the subspace with a given $\textbf{s}$ as $\hat{S}_j=\hat{\sigma}_j^z\hat{\sigma}_{j+1}^{z}$ is the $j$-th stabilizer operator. 
$\hat{R}_{\bf s}$ are the corresponding recoveries from the error subspace to the code space, with $w_{\bf s}$ the weight of the syndrome $\textbf{s}$.

In Fig.~\ref{fig:repetition_code} (a) we plot the code emission $\mathcal{E} = \sum_{\bf s}\hat{L}_{\bf s}^{\textrm{corr}\,\dagger}\hat{L}_{\bf s}^{\textrm{corr}}$ conditioned on a bit flip or on its absence [c.f. Eq.~\eqref{eqs:past_quantum_state_emission} for the cat code].
Here, $d$ plays the role of the cat size, and a large emission is associated with the occurrence of bit flips.
The inset reports the bit-flip error rate, which decreases with $d$.
In Fig.~\ref{fig:repetition_code} (b) we plot $P(n|Q)$ with $Q\in\{{\rm BF}, \neg {\rm BF}\}$ as a function of $d$ conditioned on the bit flip (blue lines) or on its absence (red lines).
In analogy with the cat qubit, the overlap between  $P(n|\textrm{BF})$ and $P(n|\neg\textrm{BF})$ decreases with $d$ consequently improving the distinguishability of the two events.

We conclude that bit flips are erasures also for this example of repetition code.
Relaxing the similarity to the cat code makes bit flips unheralded errors~\suppinfo.
Overall, this result corroborates that the output signal of locally stabilized, autonomously corrected quantum codes carries important information about the occurrence of logical errors.

\clearpage
\onecolumngrid

\begin{center}
  \textbf{\large Supplementary Information for \\ ``Bit flips are erasures in dissipative cat qubits''}\\[1em]
  Filippo Ferrari, Joachim Cohen, Vincenzo Savona, and Fabrizio Minganti
\end{center}
\vspace{1em}

\titleformat{\section}{\bfseries\centering}{\thesection.}{1em}{}
\titleformat{\subsection}{\bfseries\centering}{\thesubsection}{1em}{}
\titleformat{\subsubsection}{\centering\itshape}{\thesubsubsection}{1em}{}

\renewcommand{\thesection}{\Roman{section}}
\renewcommand{\thesubsection}{\Alph{subsection}}
\renewcommand{\thesubsubsection}{\Alph{subsection}}

\subsection{I. BIT FLIPS AND QUANTUM TRAJECTORIES}

\begin{figure*}[b]
    \centering
    \includegraphics[width=\linewidth]{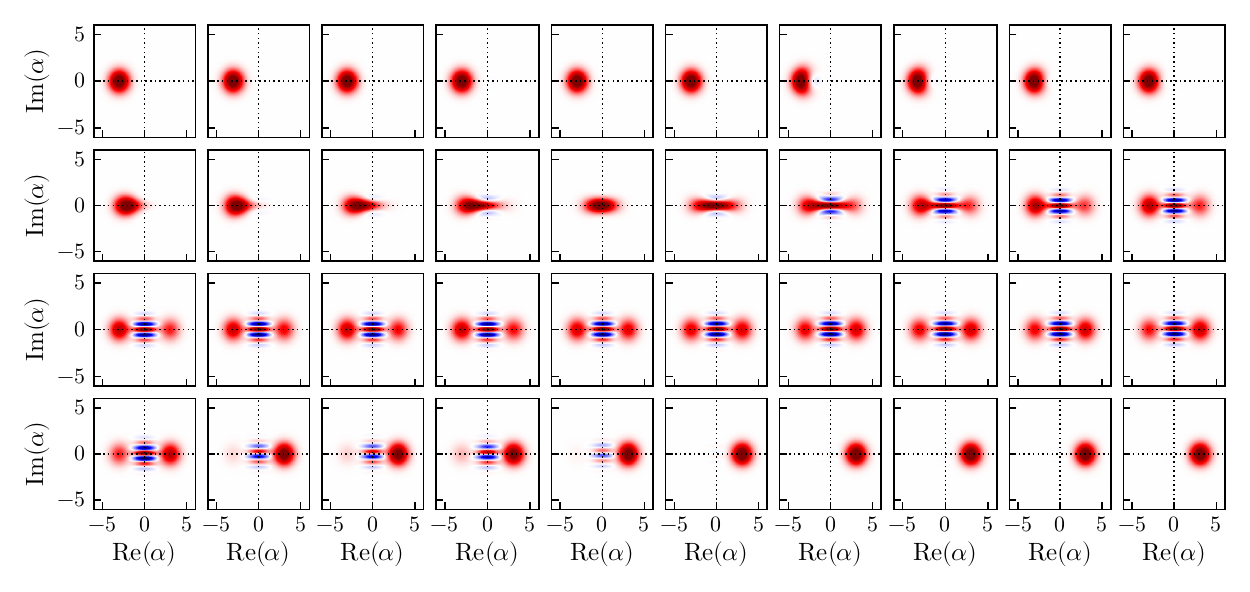}
    \caption{
    Wigner functions of the quantum trajectory $\ket{\Psi(t)}$ studied in Fig.~\ref{fig:quantum_trajectories} in the main text.
    Time flows from left to right and from the top to the bottom of the plot.
    Parameters as in Fig.~\ref{fig:quantum_trajectories} in the main text.
    }
    \label{fig:many_wigners}
\end{figure*}

In this section, we provide additional details on the bit flip dynamics within a quantum trajectory picture. 
In particular, we consider the same quantum trajectory plotted in Fig.~\ref{fig:quantum_trajectories} in the main text, and we plot the memory Wigner functions (i.e., a phase-space representation of the memory reduced density matrix~\cite{carmichael_statistical_1999, polkovnikov_phase_2010}) across the entire bit flip process in Fig.~\ref{fig:many_wigners}.
Time flows from left to right and from the top to the bottom of the plot.
Each Wigner function corresponds to a time instant across the quantum trajectory shown in Fig.~\ref{fig:many_wigners}. 
The system starts in the coherent state $\ket{\alpha}$.
It flows through a squeezed-like vacuum state.
Afterwards a cat-like state is stabilized and the single photon dissipation eventually erases the fringes.
The system ends, after the bit flip, in the coherent state $\ket{-\alpha}$.

\subsection{II. DEFINITION OF LOGICAL STATES AND PROJECTORS}

We consider an open quantum system described by a Lindblad master equation in the form

\begin{equation}
    \frac{\partial\hat{\rho}}{\partial t} = -\rmi[\hat{H},\hat{\rho}] + \sum_\mu\kappa_\mu\left(\hat{L}_\mu\hat{\rho}\hat{L}_\mu^\dagger - \frac{1}{2}\{\hat{L}_\mu^\dagger\hat{L}_\mu,\hat{\rho}\}\right) = \LL\hat{\rho},
\end{equation}
where $\hat{L}_\mu$ is a collection of Lindblad jump operators and $\LL$ is the non-Hermitian Liouvillian superoperator.
The Liouvillian can be diagonalized into a set of right and left eigenvectors, reading
\begin{equation}
    \LL\,\hat{\eta}_j = \lambda_j\,\hat{\eta}_j,\qquad\qquad
    \LL^\dagger\,\hat{\sigma}_j=\lambda_j^*\,\hat{\sigma}_j,
\end{equation}
with the biorthonormality condition $\operatorname{Tr}(\hat{\sigma}_j^\dagger\,\hat{\eta}_k) = \delta_{jk}$.
It follows that if we initialize the system in the state $\hat{\rho}(0) = \sum_jc_j\hat{\eta}_j$ with $c_j = \operatorname{Tr}[\hat{\sigma}_j^\dagger\hat{\rho}(0)]$, then $\hat{\rho}(t) = \sum_jc_j\rme^{\lambda_jt}\hat{\eta}_j$ for all times.
The same relation holds for the left eigenstates of the Liouvillian.

In the presence of a \textit{strong} $\mathbb{Z}_2$ symmetry (the same point holds also for other Liouvillian symmetries), the steady state of the system is a manifold.
By strong symmetry, we mean that the symmetry operator commutes with both the Hamiltonian and all the jump operators.
A more detailed discussion can be found, e.g., in Ref.~\cite{albert_symmetries_2014}.
A perturbative breaking of the strong symmetry due to a jump operator (for cat states, this typically coincides with single-photon losses) makes the Liouvillian symmetry \textit{weak} (now the symmetry superoperator commutes with $\mathcal{L}$, but it is no longer true that the symmetry operator commutes with all the jump operators) and the manifold metastable~\cite{macieszczak_towards_2016, macieszczak_theory_2021}.
The states and projectors associated to this metastable manifold can be derived from the Liouvillian eigenstates $\hat{\eta}_1$ and $\hat{\sigma}_1$. 
These objects encodes quantum information on the $z$ axis of the logical Bloch sphere, and the incoherent transition from the north to the south pole (a bit flip error) takes place at a time scale proportional to the Liouvillian gap $\lambda_1$.
Specifically
\begin{equation}
    \hat{\eta}_1 + \hat{\eta}_1^\dagger\propto\hat{\rho}^{+z}-\hat{\rho}^{-z},\qquad\qquad \hat{\sigma}_1 + \hat{\sigma}_1^\dagger\propto\hat{\Pi}^{+z}-\hat{\Pi}^{-z},
\end{equation}
where $\hat{\rho}^{\pm z}$ are density matrices and the projectors $\hat{\Pi}^{\pm z}$ are normalized such that $\operatorname{Tr}[\hat{\rho}^{\pm z}\hat{\Pi}^{\pm z}]=1$ and
$\operatorname{Tr}[\hat{\rho}^{\pm z}\hat{\Pi}^{\mp z}]=0$.
The logical $z$ Pauli operator is defined as $\hat{\sigma}^{z}_{L}=\hat{\Pi}^{+z}-\hat{\Pi}^{-z}$.
The logical $x$ and $y$ states and projectors can be constructed from $\hat{\eta}_j$ and $\hat{\sigma}_j$ with $j=2, 3$.
Similarly to the $z$ component we get
\begin{equation}
    \hat{\eta}_2 + \hat{\eta}_2^\dagger\propto\hat{\rho}^{+x}-\hat{\rho}^{-x},\qquad \hat{\sigma}_2 + \hat{\sigma}_2^\dagger\propto\hat{\Pi}^{+x}-\hat{\Pi}^{-x},\qquad\hat{\eta}_3 + \hat{\eta}_3^\dagger\propto\hat{\rho}^{+y}-\hat{\rho}^{-y},\qquad\hat{\sigma}_3 + \hat{\sigma}_3^\dagger\propto\hat{\Pi}^{+y}-\hat{\Pi}^{-y},
\end{equation}
and the logical $x$ and $y$ Pauli operators read $\hat{\sigma}^{x}_{L}=\hat{\Pi}^{+x}-\hat{\Pi}^{-x}$ and $\hat{\sigma}^{y}_{L}=\hat{\Pi}^{+y}-\hat{\Pi}^{-y}$ respectively.
The incoherent rotation on the equator of the Bloch sphere (a phase flip error) is encoded by $\lambda_{2, 3}\propto\langle\hat{a}^\dagger\hat{a}\rangle$ and $\lambda_{2, 3}\gg\lambda_1$.
Hence, cat qubits are paradigmatic examples of noise-biased qubits.

It is important to distinguish the logical \emph{states} $\hat{\rho}^{\pm\mu}$ from the logical \emph{projectors} $\hat{\Pi}^{\pm\mu}$: in general $\hat{\Pi}^{-z}\neq\hat{\rho}^{-z}$.
The states, built from the right eigenmodes $\hat{\eta}_j$, are density matrices ($\mathrm{Tr}[\hat{\rho}^{\pm z}]=1$) localized near $\ket{\pm\alpha}$. 
The projectors, built from the left eigenmodes $\hat{\sigma}_j$, are not states ($\mathrm{Tr}[\hat{\Pi}^{\pm z}]\neq1$) but their duals, $\mathrm{Tr}[\hat{\rho}^{\pm z}\hat{\Pi}^{\pm z}]=1$, $\mathrm{Tr}[\hat{\rho}^{\pm z}\hat{\Pi}^{\mp z}]=0$. Their meaning follows from $\hat{\sigma}^{z}_{L}=\hat{\Pi}^{+z}-\hat{\Pi}^{-z}$ being an adjoint eigenoperator, $\rme^{\mathcal{L}^{\dagger}t}\hat{\sigma}^{z}_{L}=\rme^{\lambda_{1}t}\hat{\sigma}^{z}_{L}$, so that $\mathrm{Tr}[\hat{\sigma}^{z}_{L}\hat{\rho}]\simeq\mathrm{Tr}[\hat{\sigma}^{z}_{L}\,\rme^{\mathcal{L}t}\hat{\rho}]$ for $t\ll|\lambda_{1}|^{-1}$. Choosing $t$ in the window $\Gamma_{\rm rec}^{-1},|\lambda_{2,3}|^{-1}\ll t\ll|\lambda_{1}|^{-1}$---non-empty precisely because of the noise bias $|\lambda_1|\ll|\lambda_{2,3}|\ll\Gamma_{\rm rec}$, with $\Gamma_{\rm rec}$ the confinement rate set by $\kappa_2=4g_2^2/\kappa_b$---the state relaxes onto the manifold, $e^{\mathcal{L}t}\hat{\rho}\simeq p_{+z}\hat{\rho}^{+z}+p_{-z}\hat{\rho}^{-z}$, and one finds $\mathrm{Tr}[\hat{\Pi}^{\pm z}\hat{\rho}]\simeq p_{\pm z}$. 
Thus, the projectors $\hat{\Pi}^{\pm z}$ act as ``syndrome\,$+$\,decoder'': it asks not whether the state \emph{is} the pole $\hat{\rho}^{\pm z}$, but whether it lies in that pole's basin of attraction and would be recovered there. The same holds for $x,y$ with the associated eigenvalues $\lambda_{2,3}$.

\section{III. NUMBER RESOLVED MASTER EQUATION}

Suppose we want to count the quantum jumps generated by a specific channel along a dissipative evolution by looking at the smooth, averaged time evolution generated by the Liouvillian $\LL$.
This corresponds to the full counting statistics~\cite{menczel_full_2026}.
To do this, we split the Liouvillian as $\LL = \LL_0 + \JJ$ where $\LL_0$ is the generator of the time evolution without the quantum jumps described by $\JJ$.
Given an initial state $\hat{\rho}(0)$ and a time span $[0, T]$, the probability density of obtaining the jump record $(t_1,\,...,\, t_n)$ reads
\begin{equation}
    P(t_1,\,...,\, t_n) = \operatorname{Tr}\left[\rme^{\LL_0(T-t_n)}\JJ\rme^{\LL_0(t_n-t_{n-1})}\,...\,\JJ\rme^{\LL_0t_1}\hat{\rho}(0)\right],
\end{equation}
and the state at final time, conditioned on the jump record, is 
\begin{equation}
    \hat{\rho}_{T|t_1\,...\,t_n} = \frac{\rme^{\LL_0(T-t_n)}\JJ\rme^{\LL_0(t_n-t_{n-1})}\,...\,\JJ\rme^{\LL_0t_1}\hat{\rho}(0)}{ \operatorname{Tr}\left[\rme^{\LL_0(T-t_n)}\JJ\rme^{\LL_0(t_n-t_{n-1})}\,...\,\JJ\rme^{\LL_0t_1}\hat{\rho}(0)\right]}.
\end{equation}
These conditioned states are called trajectories. 
The average over all the trajectories gives back the density matrix obtained from the standard Lindblad master equation, namely
\begin{equation}\label{eq:integral_convolution}
    \hat{\rho}(T) = \sum_n\int_0^T\rmd t_1\,...\,\int_{t_{n-1}}^T\rmd t_n\, \rme^{\LL_0(T-t_n)}\JJ\rme^{\LL_0(t_n-t_{n-1})}\,...\,\JJ\rme^{\LL_0t_1}\hat{\rho}(0)=\sum_n\hat{\rho}_n(T).
\end{equation}
The (unnormalized) states $\hat{\rho}_n$ represent the states for which, at time $T$, the jump count is $n$. 
Therefore, $P(n)=\operatorname{Tr}(\hat{\rho}_n)$ is the probability that the jump count has the value $n$ at time $T$.
To find $\hat{\rho}_n(T)$, given $\hat{\rho}(0)$ and $T$, we recognize from Eq.~\eqref{eq:integral_convolution} that the $n$-th state $\hat{\rho}_n$ can be expressed in terms of the $(n-1)$-th state as
\begin{equation}
    \hat{\rho}_n(T) = \int_0^T\rmd\tau\,\rme^{\LL_0(T-\tau)}\JJ\hat{\rho}_{n-1}(\tau).
\end{equation}
Differentiating in time the above relation and using the Leibniz integral rule yields the cascade of differential equations
\begin{equation}\label{eqs:fcs}
\begin{split}
     \frac{\partial}{\partial t}\hat{\rho}_n &= \LL_0\hat{\rho}_n + \JJ\hat{\rho}_{n-1},\\
     \frac{\partial}{\partial t}\hat{\rho}_{n-1} &= \LL_0\hat{\rho}_{n-1} + \JJ\hat{\rho}_{n-2},\\
     &\vdots \\
     \frac{\partial}{\partial t}\hat{\rho}_0 &= \LL_0\hat{\rho}_0,
\end{split}
\end{equation}
with $\hat{\rho}_0(t=0) = \hat{\rho}^{\pm\nu}$ with $\nu=\{x, y, z\}$, $\hat{\rho}_{n>0}(0) = 0$ and $t\in[0, T]$.

\begin{figure*}
    \centering
    \includegraphics[width=0.4\linewidth]{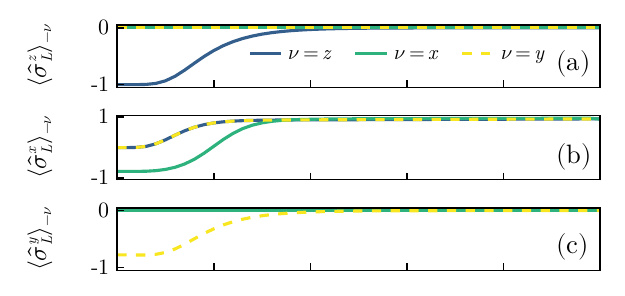}
    \caption{
    Expectation value of the logical Pauli operator, $\langle\hat{\sigma}_L^\mu(T, n)\rangle_{-\nu}$ corresponding to $\hat{\rho}(t=0)=\hat{\rho}^{-\nu}$ with $\nu=\{x, y, z\}$ (blue, green and yellow lines respectively), for (a) $\mu=z$, (b) $\mu=x$, (c) $\mu=y$, as a function of $n$.
    The memory photon number is $\langle\hat{a}^\dagger\hat{a}\rangle=4$.
    Parameters as in Fig.~\ref{fig:past_quantum_state} in the main text.
    }
    \label{fig:pauli_negative}
\end{figure*}

The numerical resolution of Eqs.~\eqref{eqs:fcs} (and Bayes' theorem) gives access to the various probability distributions considered in the main text, such as $P(n)$, $P(Q|n)$ and $P(n|Q)$, but also to the expectation values of the logical Pauli operators
$\langle \hat{\sigma}_L^{\mu}(T, n) \rangle_{\pm\nu} = \operatorname{Tr}\left[\hat{\rho}^{\pm\nu}_{n}(T)\,\hat{\sigma}_L^{\mu}\right]/\operatorname{Tr}[\hat{\rho}^{\pm\nu}_{n}(T)]$ given the detection of $n$ buffer jumps.
In the main text, we studied the behavior of $\langle \hat{\sigma}_L^{\mu}(T, n) \rangle_{+\nu}$ [see Figs.~\ref{fig:full_counting_statistics} (a-c)].
Here, we study $\langle \hat{\sigma}_L^{\mu}(T, n) \rangle_{-\nu}$, i.e., $\hat{\rho}(t=0)=\hat{\rho}^{-\nu}$.
Results are reported in Fig.~\ref{fig:pauli_negative}.
In analogy with Figs.~\ref{fig:full_counting_statistics} (a-c), upon the occurrence of a sufficient number of jumps, all initial states collapse approximately to the $\hat{\rho}^{+x}$ state.

\section{IV. NON-IDEALITIES IN THE PHOTODETECTOR AND HOMODYNE DETECTION}

The number-resolved master equation formalism, modeling an ideal photodetector, allows one to include non-idealities in the photon counting process as well.
In standard photodetectors, there are essentially two: a finite efficiency $\eta$ of the device (some photons emitted from the cavity are not detected by the device), and a nonzero dark count rate $\gamma_\rmd$ (the device detects some photons that were not emitted by the cavity).
To account for these effects in Eqs.~\eqref{eqs:fcs}, we simply need to modify $\mathcal{J}$.
For an ideal detection of the photons emitted by the buffer, we have $\mathcal{J}\hat{\rho} = \kappa_b\hat{b}\hat{\rho}\hat{b}^\dagger$.
To take into account finite detector efficiency and a nonzero dark-count rate, we have $\mathcal{J}\hat{\rho} = \eta\kappa_b\hat{b}\hat{\rho}\hat{b}^\dagger+\gamma_\rmd\hat{\rho}$.\\

Concerning the homodyne detection scheme, the formalism of past quantum states can be straightforwardly applied to the quantities considered in the End Matter, namely the homodyne photocurrent $I_{\rm hom}$, the Cauchy-Schwarz ratio $C$ and the conditioned emission spectrum $S_{\rm out}(\omega)$ (the latter has been already derived in the End Matter).

In terms of past quantum states (i.e., conditioning the homodyne photocurrent on the occurrence or absence of bit flips) we have
\begin{equation}\label{eq:integrated_homodyne}
\begin{split}
    I_{\rm hom} = \sqrt{\frac{\kappa_b}{T}}\int_0^T\rmd t\frac{\operatorname{Tr}\left[\hat{E}(t)(\hat{b} + \hat{b}^\dagger)\hat{\rho}(t)\right]}{\operatorname{Tr}\left[\hat{E}(t)\hat{\rho}(t)\right]} + \mathcal{N}(0, 1).
\end{split}
\end{equation}
For $\mathcal{C}$, we just need to express the output buffer field conditioned on the occurrence or absence of bit flips. It reads
\begin{equation}\label{eq:field}
    \langle \hat{b} \rangle(t) = \sqrt{\kappa_b}\frac{\operatorname{Tr}\left[\hat{E}(t)\,\hat{b}\,\hat{\rho}(t)\right]}{\operatorname{Tr}\left[\hat{E}(t)\hat{\rho}(t)\right]}.
\end{equation}
For both Eqs.~\eqref{eq:integrated_homodyne} and \eqref{eq:field}, $\hat{\rho}(t)$ and $\hat{E}(t)$ evolve according to Eqs.~\eqref{eqs:past_quantum_state_definition} in the main text.

\section{V. DISSIPATIVE REPETITION CODES}
In this appendix we expand the study on dissipative repetition codes presented in the End Matter.
The structure of the code consists of a single logical qubit encoded into $d$ physical qubits, each of which is a quantum two-level system described by the Pauli operators $\hat{\sigma}_j^{\mu}$, $\mu\in\{x, y, z\}$, $j=1,...,d$.
The parameter $d$ is called the code's distance.
The logical states are $\ket{0_L}=\ket{0\,...\,0}$ or $\ket{1_L}=\ket{1\,...\,1}$.
Each qubit is subject to bit-flip noise at rate $\gamma$, described by the Lindblad jump operator $\hat{L}_j^{\rm bf} = \sqrt{\gamma/2}\,\hat{\sigma}_j^x$.
Autonomous quantum error correction for bit-flip noise is implemented by a second set of jump operators which detect errors and bring the system back to the code space.
To build them up, we consider the set of stabilizer operators $\hat{S}_j = \hat{\sigma}_j^z\hat{\sigma}_{j+1}^z$, $j=1,...,d-1$.
The logical states are eigenstates of $\hat{S}_j$ with eigenvalue $+1$.
The bit-flip jump operator $\hat{L}_j^{\rm bf}$ anticommutes with $\hat{S}_{j-1}$ and $\hat{S}_j$ and flips their eigenvalues to $-1$.
Now, given a flip pattern $\mathbf{e}\in\{0, 1\}^d$ (a $1$ marks a flipped qubit), each stabilizer measures the parity of its two qubits, $s_j = (-1)^{e_j\oplus e_{j+1}}$, so that $s_j=-1$ when \emph{exactly one} of the neighboring qubits $j$ and $j+1$ is flipped, i.e., at a domain wall separating a flipped from an unflipped region.
The collection of outcomes $\mathbf{s} = (s_1,...,s_{d-1})\in\{+1,-1\}^{d-1}$ is called the \textit{syndrome}; since each of the $d-1$ stabilizers can independently take two values, there are $2^{d-1}$ distinct syndromes.
The code detects the syndrome and brings the system back to the code space.
We consider the projectors onto the subspace with a given syndrome, defined as
\begin{equation}
    \hat{P}_{\bf s} = \prod_{j=1}^{d-1}\frac{\mathds{1}+s_j\hat{S}_j}{2}.
\end{equation}
These projectors are such that $\sum_{\bf s}\hat{P}_{\bf s}=\mathds{1}$. 
Similarly, we can define the projector onto the code space as the all-$+1$ case, $\hat{P}_{\rm code} = \prod_{j=1}^{d-1}(\mathds{1}+\hat{S}_j)/2$.
Finally, for the decoded pattern $\mathbf{e}$ of a syndrome, the recovery operator is by definition a sequence of flips undoing it.
Two options are given: either we undo the flips all at once (global recovery), or we do it sequentially (sequential recovery).
This procedure is implemented by the two different recovery operators
\begin{equation}\label{eq:recovery}
    \hat{R}_{\bf s}^{\rm glob} = \prod_{j:\,e_j=1}\hat{\sigma}_j^x,\qquad\qquad\hat{R}_{\bf s}^{\rm seq} = \frac{1}{\sqrt{w_{\bf s}}}\sum_{j:\,e_j=1}\hat{\sigma}_j^x,
\end{equation}
where $w_{\bf s} = \sum_{j} e_j$ is the weight of the recovery, i.e., the number of qubits that the (minimum-weight) decoded pattern $\mathbf{e}$ flips to undo the syndrome $\mathbf{s}$ (equivalently, the number of terms in the sum).
The factor $1/\sqrt{w_{\bf s}}$ normalizes the sequential recovery so that both recoveries return the system to the code space at the same rate: since the cross terms $\hat{P}_{\bf s}\hat{\sigma}_j^x\hat{\sigma}_k^x\hat{P}_{\bf s}$ vanish for $j\neq k$ within a syndrome sector, one finds $\hat{R}_{\bf s}^{\dagger}\hat{R}_{\bf s}\hat{P}_{\bf s} = \hat{P}_{\bf s}$ for both choices in Eq.~\eqref{eq:recovery}.
The error-correcting jump operator is therefore the composition of syndrome projection $\hat{P}_{\bf s}$ and recovery $\hat{R}_{\bf s}$ at rate $\kappa\gg\gamma$,
\begin{equation}
    \hat{L}_{\bf s}^{\rm corr} = \sqrt{\kappa}\hat{R}_{\bf s}\hat{P}_{\bf s}.
\end{equation}
Since there is one error-correcting jump operator per nontrivial syndrome, their number is $2^{d-1}-1$.
In the main text, we considered the right-hand side of Eq.~\eqref{eq:recovery}, so sequential recovery.
In this section we consider both possibilities.

\begin{figure*}[t]
    \centering
    \includegraphics[width=1\linewidth]{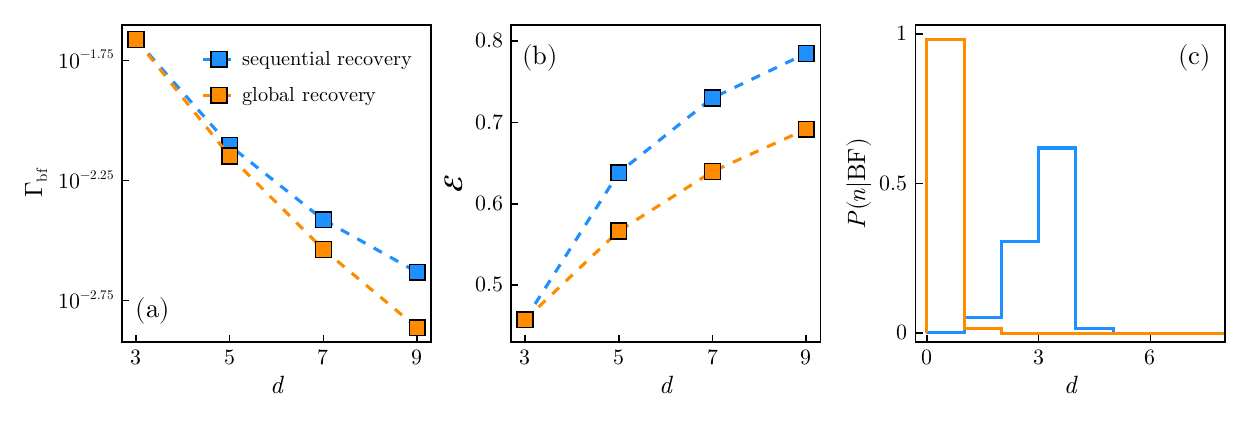}
    \caption{
    Bit flips in an autonomously stabilized repetition code.
    (a) Scaling of the bit-flip error rate $\Gamma_{\rm bf}$ with the code distance $d$ for the global (orange markers) and sequential (blue markers) recoveries in the code.
    (b) mean ``code emission" $\mathcal{E}$ given by Eqs.~\eqref{eqs:code_emission} conditioned on a bit flip as a function of $d$ for the two codes.
    (c) full counting distribution conditioned on the occurrence of a bit flip, $P(n|\textrm{BF})$ for $d=9$ and for the two codes.
    $\mathcal{E}$ and $P(n|\textrm{BF})$ are calculated for an integration time $T=\kappa$.
    The parameters of the repetition codes are $\gamma=0.1$ and $\kappa=1$.
    }
    \label{fig:repetition_code_supp}
\end{figure*}

To study the erasure nature of bit flips in this class of autonomously stabilized quantum codes, we adopt the past quantum state and number-resolved master equation formalisms developed above. 
First, we adapt the language used for dissipative cat qubits to the dissipative repetition code.
The logical states and projectors simply read
\begin{equation}
    \hat{\rho}^{+z} = \ketbra{0_L},\qquad\hat{\rho}^{-z} = \ketbra{1_L},\qquad\hat{\Pi}^{+z} = \ketbra{0_L},\qquad\hat{\Pi}^{-z} = \ketbra{1_L},
\end{equation}
and Eqs.~\eqref{eqs:past_quantum_state_definition} in the main text apply verbatim.
The conditioned ``code emission" follows from Eq.~\eqref{eqs:past_quantum_state_emission} in the main text
\begin{equation}\label{eqs:code_emission}
    n_{\rm out}(t)
    =
    \frac{
        \operatorname{Tr}\!\left[
            \hat{E}(t)\,\hat{O}_\textrm{diss}\,\hat{\rho}(t)
        \right]
    }{
        \operatorname{Tr}\!\left[
            \hat{E}(t)\hat{\rho}(t)
        \right]
    } , \qquad\qquad\mathcal{E}
    = \int_0^T \rmd t\, n_{\rm out}(t).
\end{equation}
with $\hat{O}_{\rm diss}=\sum_{\bf s}\hat{L}_{\bf s}^{\textrm{corr}\,\dagger}\hat{L}_{\bf s}^{\rm corr}$.

In Fig.~\ref{fig:repetition_code_supp} (a) we plot the bit-flip error rate $\Gamma_{\rm bf}$, i.e., the Liouvillian gap $-\textrm{Re}(\lambda_1)$, as a function of the code distance $d$ for the global (orange markers) and sequential (blue markers) recoveries in the repetition code.
Notably, compared to the cat code, $d$ plays the role of $\langle\hat{a}^\dagger\hat{a}\rangle$: $\Gamma_{\rm bf}$ shows a decrease with $d$ for both the global and sequential recoveries.
The two curves nearly coincide, as expected: by virtue of the rate normalization introduced above, the two recoveries share the same rate operator $\hat{O}_{\rm diss}=\kappa\sum_{\bf s}\hat{P}_{\bf s}=\kappa(\mathds{1}-\hat{P}_{\rm code})$, hence the same no-jump dynamics that sets the leading correction strength, so the gap is essentially insensitive to the choice of recovery.
Unlike the cat code, however, this decrease is \emph{not} exponential in $d$.
The reason is rooted in the structure of the correction.
Both recoveries act as a single, syndrome-global channel of fixed rate $\kappa$, whereas bit-flip noise is injected \emph{independently} on each of the $d$ physical qubits, so the total error influx grows as $\sim d\gamma$.
The protection is therefore governed by the ratio $d\gamma/\kappa$, whose base itself grows with the code distance: the marginal gain from each added qubit diminishes as the influx $d\gamma$ approaches the fixed correction throughput $\kappa$, and the suppression saturates rather than continuing exponentially.
By contrast, the cat code encodes $\ket{0_L}$ and $\ket{1_L}$ into the quasi-orthogonal coherent states $\ket{\pm\alpha}$, whose overlap $\sim \rme^{-2\langle\hat{a}^\dagger\hat{a}\rangle}$ is exponentially small, and each pole is stabilized \emph{independently} by the two-photon dissipation.
Increasing the cat size $\langle\hat{a}^\dagger\hat{a}\rangle$ thus raises the bit-flip barrier with a fixed base and without any shared correction channel whose throughput must scale with the code, so the suppression remains exponential.

In Fig.~\ref{fig:repetition_code_supp} (b) we plot the ``code emission" $\mathcal{E}$ conditioned on the occurrence of bit flips for the two recovery schemes.
We observe that the global recovery emits less than the sequential one, so that it heralds the bit flip more weakly.
The origin of this difference is purely dynamical: since the two recoveries share the same rate operator $\hat{O}_{\rm diss}$, the instantaneous emission rate is set by the same observable, and the gap in $\mathcal{E}$ comes entirely from how the correction proceeds in time.
The global recovery undoes all flips in a single jump, teleporting the state back to the code space ``all at once" and releasing essentially one quantum of emission.
The sequential recovery instead peels one flip per jump, so that clearing a weight-$w_{\bf s}$ error requires a cascade of up to $w_{\bf s}$ correction jumps, each contributing to the emission record.
This distinction is even more visible in the full counting distribution conditioned on the bit flip, $P(n|\textrm{BF})$, shown in Fig.~\ref{fig:repetition_code_supp} (c).
In this case, the global recovery (orange line) is associated with a distribution concentrated around $n=1$, while for the sequential recovery (blue line) $P(n|\textrm{BF})$ extends over $n\gtrsim4$, improving the distinguishability between the bit- and no-bit-flip cases.
The sequential recovery therefore grafts the erasure-heralding of an extended relaxation onto the otherwise optimal global syndrome decoding, at no cost in the protection gap.

\end{document}